\documentclass[aps,pra,showpacs,superscriptaddress,preprint]{revtex4}

\usepackage{amsmath}

\usepackage{epsf}

\usepackage{epsfig}

\usepackage{bm}

\catcode`ð=\active

 \defð{\u{g}}

 \catcode`Ð=\active

 \defÐ{\u{G}}

 \catcode`Ý=\active

 \defÝ{\. I}

 \catcode`ö=\active

 \defö{\"{o}}

 \catcode`Ö=\active

 \defÖ{\"O}

 \catcode`ü=\active

 \defü{\"{u}}

 \catcode`Ü=\active

 \defÜ{\"{U}}

 \catcode`Þ=\active

 \defÞ{\c{S}}

 \catcode`þ=\active

 \defþ{\c{s}}

 \catcode`ý=\active

 \defý{{\i}}

 \catcode`ç=\active

 \defç{\d{c}}

 \catcode`Ç=\active

 \defÇ{\d{C}}

\begin{document}

\title{Bound State Solutions of the Schrödinger Equation for Generalized
Morse Potential With Position Dependent Mass}

\author{\small Altuð Arda}
\email[E-mail: ]{arda@hacettepe.edu.tr}\affiliation{Department of
Physics Education, Hacettepe University, 06800, Ankara,Turkey}
\author{Ramazan Sever}
\email[E-mail: ]{sever@metu.edu.tr}\affiliation{Department of
Physics, Middle East Technical University, 06800, Ankara,Turkey}

\begin{abstract}
The effective mass one-dimensional Schrödinger equation for the
generalized Morse potential is solved by using Nikiforov-Uvarov
method. Energy eigenvalues and corresponding eigenfunctions are
computed analytically. The results are also reduced to the case of
constant mass. Energy eigenvalues are computed numerically
for some diatomic molecules. The results are in agreement with the ones obtained before.
\end{abstract}

\maketitle

\newpage

\section{Introduction}
In recent years, the effects of the coordinate-dependence of the
mass on the solutions of the relativistic and/or non-relativistic
wave equations have been received a great attentions[1-14].This is
so because it has a widely range applications on different areas,
for example, in the study of the semiconductors [15], or of
electronic properties of quantum wells and quantum dots [16], and
the impurities in crystals [17-19]. Point canonical transformation
[9], deformed algebras [20], quadratic algebra method [21], path
integral method [22], Lie algebra approach [23-25],
group-theoretical approach [26], and supersymmetric formalism [27]
are some of the methods used in the literature to solve the wave
equations for the case of constant and/or position-dependent mass
(PDM) distributions.

In the present work, we study the effects of the PDM on the
solutions of the Schrödinger equation (SE) for the generalized
Morse potential given by

\begin{eqnarray}
V(x)=V_1e^{-2\beta\,(\frac{r-r_0}{r_0})}-V_2e^{-\beta\,(\frac{r-r_0}{r_0})}\,,
\end{eqnarray}
where $V_1$\,, and $V_2$ are two parameters, and correspond to
$D$, and $2D$ in the usual Morse potential, respectively, where
$D$ is the dissociation energy of the molecule, and the parameter
$\beta$ is the depth of the potential. The Morse potential is one
of the important exponentially varying potential which describes
the interaction in diatomic molecules [28]. The Morse potential is
exactly solvable for $s$-waves. However, for any $\ell$-states
only numerical solutions can be obtained with different
approximation techniques [29]. We intend to apply the
Nikiforov-Uvarov (NU) method [30] to find out the energy spectra
and corresponding eigenfunctions of the generalized Morse
potential by using a suitable coordinate-dependence mass
distribution function which is finite at infinity, and also
enables us to solve the SE exactly.

The work is organized as follows. In Section II, we find out the
energy eigenvalues and corresponding eigenfunctions of the SE by
using the NU-method in the case of PDM. We give also the results
for the case of constant mass. We give the numerical results for
the bound state energies of $H_2$, and $LiH$ molecules in the case
of constant and spatially dependent mass in Table I. We summarize
our conclusions in Section III.

\section{NIKIFOROV-UVAROV METHOD and BOUND STATES}

The one-dimensional time-independent effective mass SE equation
with the potential $V(x)$ reads [6]

\begin{eqnarray}
-\,\frac{\hbar^2}{2m(x)}\,\frac{d^2\psi(x)}{dx^2}+\,\frac{\hbar^2}{2}\,\Bigg(\frac{m^{\prime}(x)}{m^2(x)}\Bigg)
\,\frac{d\psi(x)}{dx}+\Big[V(x)+U_{\alpha\beta\gamma}(x)-E\Big]\psi(x)=0,
\end{eqnarray}
where prime denotes the derivative of mass $m(x)$ with respect to
position and

\begin{eqnarray}
U_{\alpha\beta\gamma}(x)=-\,\frac{\hbar^2}{4m^{3}(x)(a+1)}
\Bigg[(\alpha+\gamma-a)m(x)m^{\prime\prime}(x)+2(a-\alpha\gamma-\alpha-\gamma)m^{\prime2}(x)\Bigg].
\end{eqnarray}
where the ambiguity parameters proposed by different authors in
the literature satisfy $\alpha+\beta+\gamma=-1$. In the ordering
proposed by Weyl, the parameters have the values $a=1$,
$\alpha=\gamma=0$. In the parameter set due to Li and Kuhn
$a=\alpha=0$, and $\gamma=-\,\frac{1}{2}$, etc. [6]. It is found
in Ref. [6] that the Weyl and Li and Kuhn ambiguity orderings are
equivalent in the case of SE equation.

By redefinition of the wave function

\begin{eqnarray}
\psi(x)=\sqrt{m(x)}\phi(x),
\end{eqnarray}
we get the SE as

\begin{eqnarray}
-\frac{\hbar^2}{2m(x)}\,\frac{d^2\phi(x)}{dx^2}+(U_{eff}(x)-E)\phi(x)=0.
\end{eqnarray}
The effective potential in the above equation is

\begin{eqnarray}
U_{eff}(x)=U_{\alpha\beta\gamma}(x)+V(x)+\,\frac{\hbar^2}{4m^{2}(x)}\Bigg(\frac{3m^{\prime2}(x)}{2m(x)}\,
-m^{\prime\prime}\Bigg).
\end{eqnarray}

We consider the generalized Morse potential, which can be used to
describe the vibrations of a two-atomic molecule, as the following

\begin{eqnarray}
V(x)=V_1 e^{-2\beta x}-V_2e^{-\beta x}\,,\,\,(0 \leq x \leq
\infty)\,.
\end{eqnarray}
where $V_1$, and $V_2$ are two real parameters, and $x=(r-r_0)$.
The parameter $\beta$ is $\alpha'/r_0$, here, $\alpha'$ is a
positive parameter, and $r_0$ is the equilibrium distance [29].

Here, we prefer to use the following mass-distribution

\begin{eqnarray}
m(x)=\,\frac{m_0}{(1-\eta e^{-\beta x})^2}\,,
\end{eqnarray}
where $m_0$ corresponds to the constant mass, and the free
parameter $\eta$ satisfies the condition that $\eta < 1$ to handle
a positive mass at the zero. We get a finite mass distribution
with this constraint on $\eta$. We can check out the results of
the case of constant mass. Further, we solve the SE analytically
by using such a mass-function.

Substituting Eqs. (8) and (7) into Eq. (5) we get

\begin{eqnarray}
\frac{d^2\phi(x)}{dx^2}\,&+&\,\frac{1}{(1-\eta e^{-\beta
x})^2}\,\Big\{\,\Big[2A_2 \beta^2 \eta +\,\frac{2m_0
V_2}{\hbar^2}\Big]e^{-\beta x}\nonumber\\&+&\Big[\,4\beta^2
\eta^2(A_1+A_2)-\,\frac{2m_0 V_1}{\hbar^2}\Big]e^{-2\beta
x}+\,\frac{2m_0 E}{\hbar^2}\Big\}\,\phi(x)=0\,,
\end{eqnarray}
where

\begin{eqnarray}
A_1&=&\,\frac{a-\alpha\gamma-\alpha-\gamma}{1+a}\,-\,\frac{3}{4}\,,\nonumber\\
A_2&=&\,\frac{\alpha+\gamma-a}{2(1+a)}\,+\,\frac{1}{2}\,.
\end{eqnarray}

By using the transformation $z=e^{-\beta x}\,\,(0 \leq z \leq 1)$,
we obtain

\begin{eqnarray}
\frac{d^2\phi(z)}{dz^2}\,+\,\frac{1-\eta z}{z(1-\eta
z)}\,\frac{d\phi(z)}{dz}\,+\,\frac{1}{[\eta(1-\eta z)
]^2}\Big(-\varepsilon_1 z^2-\varepsilon_2
z-\epsilon_{n\ell}\Big)\phi(z)=0\,,
\end{eqnarray}
where

\begin{eqnarray}
-\varepsilon_1&=&4\eta^2\Big(\,\frac{a-2\alpha\gamma-\alpha-\gamma}{2(1+a)}\,-\,\frac{1}{4}\Big)-\,\frac{2m_0
V_1}{\beta^2\hbar^2}\,,\nonumber\\
-\varepsilon_2&=&\eta\Big(\,\frac{\alpha+\gamma+1}{1+a}\Big)+\,\frac{2m_0
V_2}{\beta^2\hbar^2}\,,\nonumber\\
-\epsilon_{n\ell}&=&\,\frac{2m_0E}{\beta^2\hbar^2}\,.
\end{eqnarray}

Now to apply the NU-method [30], we rewrite Eq. (11) in the
following form

\begin{eqnarray}
\phi^{\prime\prime}(z)+\,\frac{\tilde{\tau}(z)}{\sigma(z)}\,\phi^{\prime}(z)
+\,\frac{\tilde{\sigma}(z)}{\sigma^2(z)}\,\phi(z)=0,
\end{eqnarray}
where $\sigma(z)$ and $\tilde{\sigma}(z)$ are polynomials with
second-degree, at most, and $\tilde{\tau}(z)$ is a polynomial with
first-degree. We define the total wave function as

\begin{eqnarray}
\phi(z)=\xi(z)\psi(z).
\end{eqnarray}

Thus Eq. (13) is reduced to a hypergeometric type equation

\begin{eqnarray}
\sigma(z)\psi^{\prime\prime}(z)+\tau(z)\psi^{\prime}(z)+\lambda\psi(z)=0.
\end{eqnarray}

We also define the new eigenvalue for the Eq. (13) as

\begin{eqnarray}
\lambda&=&\lambda_n=-n\tau^{\prime}-\,\frac{n(n-1)}{2}\,\sigma^{\prime\prime}\,,
(n=0, 1, 2, \ldots)
\end{eqnarray}
where

\begin{eqnarray}
\tau(z)&=&\tilde{\tau}(z)+2\pi(z).
\end{eqnarray}

The derivative of $\tau(z)$ must be negative. $\lambda(\lambda_n)$
is obtained from a particular solution of the polynomial
$\psi_n(z)$ with the degree of $n$. $\psi_n(z)$ is the
hypergeometric type function whose solutions are given by [30]

\begin{eqnarray}
\psi_n(z)=
\,\frac{b_n}{\rho(z)}\,\frac{d^n}{dy^n}[\sigma^n(z)\rho(z)],
\end{eqnarray}
where the weight function $\rho(z)$ satisfies the equation

\begin{eqnarray}
\frac{d}{dz}[\sigma(z)\rho(z)]=\tau(z)\rho(z).
\end{eqnarray}

On the other hand, the function $\xi(z)$ satisfies the relation

\begin{eqnarray}
\xi^{\prime}(z)/\xi(z)=\pi(z)/\sigma(z).
\end{eqnarray}

Comparing Eq. (11) with Eq. (13), we have

\begin{eqnarray}
\tilde{\tau}(z)=1-\eta z\,,\,\,\,\,\,\sigma(z)=z(1-\eta
z)\,,\,\,\,\,\,
\tilde{\sigma}(z)=-\varepsilon_1z^2-\varepsilon_2z-\epsilon_{n\ell}
\end{eqnarray}

The $\pi(z)$ has the form [30]

\begin{eqnarray}
\pi(z)=\,\frac{\sigma^{\prime}(z)-\tilde{\tau}(z)}{2}\,
\pm\,\sqrt{(\frac{\sigma^{\prime}(z)-\tilde{\tau}(z)}{2})^2-\tilde{\sigma}(z)+k\sigma(z)}\,,
\end{eqnarray}
or, explicitly

\begin{eqnarray}
\pi(z)=\,-\,\frac{\eta
z}{2}\,\pm\sqrt{\,\Big(\,\frac{\eta^2}{4}\,-k
\eta+\varepsilon_1\Big)z^2+(\varepsilon_2+k)z
+\epsilon_{n\ell}}\,.
\end{eqnarray}

The constant $k$ is determined by imposing a condition such that
the discriminant under the square root should be zero. The roots
of $k$ are $k_{1,2}=-\varepsilon_2-2\eta
\epsilon_{n\ell}\mp\sqrt{\epsilon_{n\ell}}A$, where
$A=\sqrt{4\eta^2\epsilon_{n\ell}+4\eta\varepsilon_2+\eta^2+4\varepsilon_1}$.

Substituting these values into Eq. (23), we get for $k_1$

\begin{eqnarray}
\pi(z)=\,-\,\frac{\eta
z}{2}\,\mp\Big[\Big(\,\frac{A}{2}\,-\eta\sqrt{\epsilon_{n\ell}}\Big)z+\sqrt{\epsilon_{n\ell}}\Big]\,,
\end{eqnarray}
and for $k_2$

\begin{eqnarray}
\pi(z)=\,-\,\frac{\eta
z}{2}\,\mp\Big[\Big(\,\frac{A}{2}\,+\eta\sqrt{\epsilon_{n\ell}}\Big)z-\sqrt{\epsilon_{n\ell}}\Big]\,.
\end{eqnarray}

Now we calculate the polynomial $\tau(z)$  from $\pi(z)$ such that
its derivative with respect to $z$ must be negative. Thus we take
the root $k_2$, and by using Eq. (25) for $\pi(z)$ we get

\begin{eqnarray}
\tau(z)=1+2\sqrt{\epsilon_{n\ell}}-2\Big(\,\frac{A}{2}\,+\eta\sqrt{\epsilon_{n\ell}}+\eta\Big)z\,.
\end{eqnarray}
with the derivative
$\tau'(z)=-2\Big(\,\frac{A}{2}\,+\eta\sqrt{\epsilon_{n\ell}}+\eta\Big)$.

The constant $\lambda=k+\pi^{\prime}(z)$ becomes

\begin{eqnarray}
\lambda=-\varepsilon_2-2\eta\epsilon_{n\ell}-\sqrt{\epsilon_{n\ell}}\,(A+\eta)-\,\frac{1}{2}\,(A+\eta)\,,
\end{eqnarray}
and Eq. (16) gives us

\begin{eqnarray}
\lambda_n=2n\Big(\,\frac{A}{2}\,+\eta\sqrt{\epsilon_{n\ell}}+\eta\Big)+\eta
n(n-1)\,.
\end{eqnarray}

Substituting the values of the parameters given by Eq. (12), and
setting $\lambda=\lambda_n$, one can find the energy eigenvalues
as

\begin{eqnarray}
\epsilon_{n\ell}&=&\,\Bigg\{\,\frac{[n^2+n-(1/2)]\eta-\varepsilon_2-2[n+(1/2)]\sqrt{\varepsilon_1-(\eta^2/2)}}
{(2n+1)\eta -\sqrt{4\varepsilon_1-2\eta^2}}\,\Bigg\}^2\,.
\end{eqnarray}

It is seen that there is a strong dependence of the energy
eigenvalues to the parameter $\eta$. Further, one can see that the
generalized Morse potential has a real energy spectra in the case
of PDM under the condition that

\begin{eqnarray}
\frac{m_0V_1}{\beta^2\hbar^2}>\eta^2\Big(\,\frac{a-2\alpha\gamma-\alpha-\gamma}{1+a}\,-\,\frac{1}{4}\,\Big)\,.
\end{eqnarray}

We get the following energy spectra in constant mass case as

\begin{eqnarray}
\epsilon^{\eta=0}_{n\ell}=\,\frac{1}{4}\,\Big[2n+1+\,\frac{\varepsilon_2}{\sqrt{\varepsilon_1}}\Big]^2\,.
\end{eqnarray}
It is exactly same result obtained in the literature [29].

We list the numerical results for the bound state energies of the
$H_2$, and $LiH$ molecules in the constant mass case in Table I.
We use the same parameters given in Ref. (29), such as $D$, $r_0$,
$m_0$, $\alpha'$, and $E_0$, to compare our results. Here, the new
parameter $E_0$ is a short notation, i.e., $\hbar^2/(m_0r^2_0)$.
We also give the bound state energies of the above molecules for
three different values of $\eta$ for each molecule in Table I. For
simplicity, we choose the Weyl ordering for the ambiguity
parameters in the numerical analyze.

Now let us find the eigenfunctions. We first compute the weight
function from Eqs. (17) and (19)

\begin{eqnarray}
\rho(z)=z^{-2\sqrt{\epsilon_{n\ell}}}\,(1-\eta z)^{\tilde{A}}\,,
\end{eqnarray}
where
$\tilde{A}=\sqrt{1+4\epsilon_{n\ell}+\,\frac{4}{\eta}\,(\varepsilon_2+\,\frac{\varepsilon_1}{\eta})}$,
and the wave functions become

\begin{eqnarray}
\psi_n(z)=\,\frac{b_n}{z^{-2\sqrt{\epsilon_{n\ell}}}\,(1-\eta
z)^{\tilde{A}}}\,\frac{d^n}{dz^n}\,\left[
\,z^{n-2\sqrt{\epsilon_{n\ell}}}\,(1-\eta
z)^{n+\tilde{A}}\right]\,.
\end{eqnarray}
where $b_n$ is a normalization constant. In the limit $\eta
\rightarrow 1$, the polynomial solutions can be written in terms
of the Jacobi polynomials with weight function $\rho(z)$ as [31]

\begin{eqnarray}
\psi_n(z) \simeq P^{(\tilde{A},
-2\sqrt{\epsilon_{n\ell}}\,)}_n\,(2\eta
z-1)\,,\,\,\,\,\,\,\tilde{A}>-1\,,\,\,\,-2\sqrt{\epsilon_{n\ell}}\,>-1\,.
\end{eqnarray}

On the other hand, the other part of the wave function is obtained
from the Eq. (20) as

\begin{eqnarray}
\xi(z)=z^{-\sqrt{\epsilon_{n\ell}}}\,(1-\eta
z)^{(1/2)(1+\tilde{A})}\,.
\end{eqnarray}

Thus, the total eigenfunctions take

\begin{eqnarray}
\phi_n(z)=b'_n\,z^{-\sqrt{\epsilon_{n\ell}}}\,(1-\eta
z)^{(1/2)(1+\tilde{A})}P^{(\tilde{A},
-2\sqrt{\epsilon_{n\ell}}\,)}_n\,(2\eta z-1)\,.
\end{eqnarray}
where $b'_n$ is the new normalization constant. The normalization
condition $\int_{0}^{1}\,|\phi(z)|^2dz=1$ gives us

\begin{eqnarray}
b^{'2}_{n}\bigg(\frac{1}{2}\bigg)^{\tilde{A}-2\sqrt{\epsilon_{n\ell}\,}}
\int_{-1}^{+1}(1-x)^{1+\tilde{A}}(1+x)^{-2\sqrt{\epsilon_{n\ell}\,}}
P^{(\tilde{A},
-2\sqrt{\epsilon_{n\ell}}\,)}_{n}\,(x)P^{(\tilde{A},
-2\sqrt{\epsilon_{n\ell}}\,)}_{m}\,(x)dx=1\,,
\end{eqnarray}
where we use a new variable defined as $x=2\eta z-1$\,. By using
the required identities of the Jacobi polynomials [32, 33], we
obtain

\begin{eqnarray}
b^{'2}_{n}&=&\frac{2^{2\tilde{A}-4\sqrt{\epsilon_{n\ell}\,}\,+1}}{\tilde{A}-2\sqrt{\epsilon_{n\ell}\,}
+2n+1}\frac{2(\tilde{A}-\sqrt{\epsilon_{n\ell}\,})(1-\sqrt{\epsilon_{n\ell}\,}+2n)+4n(1+n)}
{(\tilde{A}-2\sqrt{\epsilon_{n\ell}\,}+2n+2)(\tilde{A}-2\sqrt{\epsilon_{n\ell}\,}+2n)}
\nonumber\\&\times&\frac{\Gamma(\tilde{A}+n+1)\Gamma(-2\sqrt{\epsilon_{n\ell}\,}+n+1)}
{n!\Gamma(\tilde{A}-2\sqrt{\epsilon_{n\ell}\,}+n+1)}\,.
\end{eqnarray}

\section{Conclusion}
We have solved the one-dimensional position-dependent effective
mass SE for the generalized Morse potential by using NU-method and
obtained analytically the energy eigenvalues and corresponding
eigenfunctions. They depend on the free parameter $\eta$ strongly.
We have shown that the results can be reduced to the ones obtained
for the case of the constant mass. We have listed the numerical
values of the energy eigenvalues in Table I for the $H_2$, and
$LiH$ molecules for different values of the quantum number $n$,
and free parameter $\eta$.

\section{Acknowledgments}
This research was partially supported by the Scientific and
Technical Research Council of Turkey.

\newpage

\begin{table}
\caption{\label{tab:special}The dependence of the bound states on
$n$ in $eV$ for $H_2$ ($D=4,7446 eV$, $r_0=0,7416 \AA$,
$m_0=0,50391 amu$, $\alpha'=1,440558$, and
$E_0=1,508343932\times10^{-2} eV$), and $LiH$ molecules
($D=2,515287 eV$, $r_0=1,5956 \AA$, $m_0=0,8801221 amu$,
$\alpha'=1,7998368$, and $E_0=1,865528199\times10^{-3} eV$) [29].}
\begin{ruledtabular}
\begin{tabular}{ccccccc}
&  &  & & &  & $\eta=0$\\ \hline
$n$ & $E_n(H_2)\footnotemark[1]$ & $E_n(H_2)\footnotemark[2]$ & & & $E_n(LiH)$\footnotemark[1] & $E_n(LiH)
\footnotemark[2]$\\
0 & -4.476 & -4.476 & & & -2.429 & -2.429\\
2 & -3.480 & -3.480 & & & -2.098 & -2.098\\
4 & -2.609 & -2.609 & & & -1.792 & -1.792\\
10 & -0.748 & -0.748 & & & -1.018 & -1.018\\
15 & -0.057 &  & & & -0.539\\
20 &  &  & & & -0.211 & -0.211\\
\hline
&  &  &  & & & $\eta\neq0$\\
\hline $n$ & $E_n(H_2)\footnotemark[3]$ &
$E_n(H_2)\footnotemark[4]$ & $E_n(H_2)\footnotemark[5]$ &
$E_n(LiH)\footnotemark[3]$
& $E_n(LiH)\footnotemark[4]$ & $E_n(LiH)\footnotemark[5]$\\
0 & -4.528 & -4.582 & -4.637 & -2.446 & -2.463 & -2.481\\
2 & -3.706 & -3.955 & -4.228 & -2.176 & -2.259 & -2.346\\
4 & -2.953 & -3.363 & -3.856 & -1.920 & -2.062 & -2.219\\
6 & -2.274 & -2.809 & -3.522 & -1.677 & -1.872 & -2.099\\
10 & -1.152 & -1.818 & -2.985 & -1.233 & -1.512 & -1.880\\
15 & -0.251 & -0.824 & -2.644 & -0.763 & -1.105 & -1.653\\
20 & -0.012 & -0.169 &  & -0.395 & -0.748 & -1.486\\
\end{tabular}
\end{ruledtabular}
\footnotetext[1]{our results} \footnotetext[2]{results obtained in
Ref [29]} \footnotetext[3]{results for $\eta=0.2$}
\footnotetext[4]{results for $\eta=0.4$} \footnotetext[5]{results
for $\eta=0.6$}
\end{table}

\end{document}